\documentclass[12pt]{article}
\usepackage{amssymb}\usepackage{a4}
\newcommand{\beao}{\begin{eqnarray*}}
\newcommand{\eeao}{\end{eqnarray*}}
\newcommand{\be}{\begin{equation}}\newcommand{\ee}{\end{equation}}
\newcommand{\bea}{\begin{eqnarray}}
\newcommand{\eea}{\end{eqnarray}}
\newcommand{\beq}{\begin{eqnarray}}
\newcommand{\eeq}{\end{eqnarray}}
\newcommand{\nn}{\nonumber}
\newcommand{\pa}{\partial}
\newcommand{\ep}{\epsilon}
\newcommand{\ga}{\gamma}
\newcommand{\om}{\omega}\newcommand{\Om}{\Omega}
\newcommand{\al}{\alpha}
\newcommand{\Ref}[1]{(\ref{#1})}
\newcommand{\F}{{\cal F}}
\renewcommand{\S}{{\cal S}}

\newcommand{\TE}{{\rm TE}}\newcommand{\TM}{{\rm TM}}
\renewcommand{\kappa}{\varkappa}
\usepackage{graphicx}

\begin{document}

\title{Drude model and Lifshitz formula}
\author{M. Bordag\footnote{bordag@itp.uni-leipzig.de}\\\small
Universit\"{a}t Leipzig, Institute for Theoretical Physics\\\small
 Box 100 920, 04009 Leipzig, Germany }
\date{}

\maketitle

\begin{abstract}
Since nearly 10 years, it is known that inserting the permittivity of the Drude model into the Lifshitz formula for  free energy causes a violation of the third law of thermodynamics. In this paper we show that the standard Matsubara formulation for  free energy contains a contribution that is non-perturbative in the relaxation parameter. We argue that the correct formula must  have a perturbative expansion and conclude that the standard Matsubara formulation with the permittivity of the Drude model inserted is not correct. We trace the non-perturbative contribution in the complex frequency plane, where it shows up as a self-intersection or a bifurcation of the integration path.\\
PACS{ {03.70.+k,}{11.10.Wx,} {65.40.gd}}
\end{abstract}


\section{Introduction}
Since nearly 10 years, we observe a problem with thermodynamics in the theory of the Casimir effect. This problem appears when calculating the free energy $\F$ of the electromagnetic field using the Lifshitz formula in the presence of metallic bodies described by the Drude model. The same happens for dielectric bodies with nonzero dc conductivity. According to the third law of thermodynamics (Nernst heat theorem), one must expect the entropy
\be\label{S}    \S=-\frac{\pa\F}{\pa T}
\ee
to vanish for $T\to0$. In fact, with the Drude permittivity\footnote{For $\ga=0$ it turns into the permittivity of the plasma model.}
\be\label{epD}\ep^{\rm D}=1-\frac{\Om^2}{\om(\om +i \ga)}
\ee
with plasma frequency $\Om$ and a relaxation parameter, $\ga(T)$, vanishing for $T\to
0$, the entropy takes a nonzero limit,
\be\label{S0}\lim_{T\to0}S=S_0,
\ee
which constitutes the violation of the third law of thermodynamics. This was found for plane parallel interfaces in \cite{beze02-66-062112,beze04-69-022119}. A similar violation was found for dielectrics with finite dc conductivity in  \cite{geye05-72-085009}. Recently, a similar finding was made for a sphere in front of a plane  \cite{bord10-82-125016}.

It must be mentioned that in a number of papers,
\cite{bost04-339-53,hoye07-75-051127,dalv08-101-163203,intr08-41-164018},
 nonzero $S_0$ is not considered as a violation of thermodynamics or it is attributed to
a perfectly symmetric crystal lattice with no impurities, which is not realized in nature \cite{milt11-79-697}. However, these discussions are not the topic of the present paper.

This violation constitutes not only a serious theoretical problem. Calculating the free energy as described above, anomalous large thermal contributions at short separation appear first predicted in \cite{bost00-84-4757}.
In a very careful short-separation experiment with metal test bodies
\cite{decc07-75-077101,decc07-51-963} these have been shown to be excluded
on a high confidence level. Similar large thermal corrections arising
from accounting dc conductivity of dielectrics were also experimentally
excluded \cite{chen07-15-4823,chen07-76-035338}.
In a large-separation experiment with centimeter-sized metal bodies \cite{masu09-102-171101} the predictions following from the use of the Drude permittivity were not supported. In line with this, it must be mentioned that in another large-separation experiment \cite{sush11-7-230} the opposite result, supporting thermal corrections predicted by the Drude permittivity, was claimed\footnote{The method used in \cite{sush11-7-230} was questioned in \cite{beze11-83-075417}.}.

In the present paper we reconsider the foundation of the theory of the Casimir effect at finite temperature and we show in which place the standard approach with the Drude permittivity fails. We start from reminding the basic formulas of this approach. In application to two dielectric half-spaces with permittivity $\ep$
separated by an empty gap with plane parallel surfaces the free energy can be expressed by the Lifshitz formula,
\be\label{F1}\F=T{\sum\limits_{l=0}^{\infty}}{\vphantom{\sum}}^{\prime}
\int\frac{d^2k_{||}}{(2\pi)^2}
\sum_{i={\rm TE,\, TM}} \ln\left(1-r_i^2\,e^{-2\eta L}\right),
\ee
where we have put $\hbar=c=k_{\rm B}=1$. The prime on the summation sign means that the term with $l=0$ has to be multiplied by 1/2.
In \Ref{F1}, the $l$-summation goes over the Matsubara frequencies,
\be\label{xil}\xi_l=2\pi lT,
\ee
and the sum over $i$ accounts for the two polarizations of the electromagnetic field. The reflection coefficients are
\bea r_{\TE}&=&\frac{\eta-\sqrt{(\ep-1)\xi_l^2+\eta^2}}{\eta+\sqrt{(\ep-1)\xi_l^2+\eta^2}},
\nn\\
r_{\TM}&=&\frac{\ep \eta-\sqrt{(\ep-1)\xi_l^2+\eta^2}}{\ep \eta+\sqrt{(\ep-1)\xi_l^2+\eta^2}},
\label{rTEM}
\eea
and the  notation
\be\label{eta}\eta=\sqrt{\xi_l^2+k_{||}^2}
\ee
is used, where $k_{||}$ is the momentum in the translational invariant directions in parallel to the interfaces. For $\ep$ one has to insert an expression according to the model chosen. If $\ep$ depends on the frequency $\om$ like \Ref{epD} does, one needs to insert $\om=i\xi_l$. This follows since the Matsubara representation \Ref{F1} assumes a summation over the imaginary frequencies.

It should be mentioned that the free energy \Ref{F1} with the Drude model permittivity, $\ep^{\rm D}$, eq. \Ref{epD}, inserted is real. This follows because on the imaginary frequency axis, the permittivity $\ep^{\rm D}$ is real unlike as on the  real frequency axis. Equation \Ref{F1} is one of a number of various equivalent representations of the Lifshitz formula. These may differ by a change of variables or by integration by parts.
This formula was initially derived more than 50 years ago \cite{lifs56-2-73}. At finite temperature, it uses the common representation of a quantum field theory at finite temperature after Matsubara. Looking at the history and on the many applications, there seems to be no reason for any doubt in the validity of \Ref{F1}. There might be only a question about the permittivity that enters through the reflection coefficients. The basic idea is that \Ref{F1} is the free energy of the electromagnetic field interacting with the dielectric bodies. This interaction is accounted for by the permittivities in the sense of  macroscopic electrodynamics.  Now, the permittivities can be taken from some theoretical model or from measurements, as well. Especially, the latter are well founded by a huge amount of data. In this way, inserting the  Drude model permittivity should give a reasonable result.

However, there is the much discussed question on what about the dissipation of energy inherent to the Drude model. This dissipation is due to  physical processes like scattering and transfer of  energy from the electromagnetic field into heat. In general,  dissipation makes the considered model incomplete as long as the return of this heat to the electromagnetic field is not accounted for or a heat bath is used like in the Huttner-Barnett type of approach \cite{hutt92-46-4306}. In physics, this return happens in a natural way since the heat is radiated back.

In general, in the quantum field theory, it is common that the energy or the free energy acquires an imaginary part if the system is not closed and looses energy, by creation of particles for example. So, if not including the return of energy, one should expect the free energy to have an imaginary part. The Lifshitz formula \Ref{F1} with permittivity \Ref{epD}, interpreted as or derived from the vacuum energy, accounts only for the degrees of freedom of the electromagnetic field in the presence of the lossy dielectrics. By this setup, there is no return of energy included. As a consequence, one should expect the free energy to have an imaginary part. It is a merit of the formula \Ref{F1} to have none.

The question on how to include dissipation was discussed intensively in \cite{bara73-16-836,bara75-18-305} (see also the explanation in \cite{milonni}, section 7.3). The main idea is to include an auxiliary field, which acts like a fluctuating source  for the electromagnetic field and returns, in this way, the dissipated energy. With some restrictions on the parameters, eq. \Ref{F1} was confirmed also for the case of dissipation \cite{bara73-16-836,bara75-18-305}. This discussion is generally considered as giving the correct result.

In our discussion, we go a similar way by dividing the system into two. The first is the electromagnetic field in the presence of the dielectric described by the Drude model. The second is some mechanism that performs the return of the dissipated energy. Then, as an approximation, we consider the first part only. Of course, this procedure is quite unsatisfactory insofar as we do not have any information on how good this approximation is. Nevertheless we will be able to gain some information on representation \Ref{F1}. In dividing the system,  we will have two contributions to the free energy. Now, both will have imaginary parts that must compensate in the sum. In this sense, it is natural to have  free energy with an imaginary part even in equilibrium if considering the first part alone.

In line with this setup, we make a point that the free energy must be perturbative in the relaxation parameter $\ga$, at least in the sense that for $\ga\to0$, the free energy turns into that of the plasma model. This follows from physical reasons in terms of the expectation that, for sufficiently small dissipation, the system should be described by the plasma model permittivity. We argue that in the underlying physics, there is no process that could create a nonvanishing contribution for vanishing dissipation. In the following, we will use the notion 'perturbative' in a generalized meaning by allowing for  expansion terms like $\ga\ln\ga$. The point is that all contributions beyond the plasma model are assumed to vanish for $\ga\to0$.

Below, we will see that the first part of the system  is indeed analytic in $\ga$.  It is interesting to mention that the representation \Ref{F1} is nonanalytic in $\ga$. This can be inferred already from the  known   observation that for $T\to 0$ the limiting value $S_0$ of the entropy does not depend on the details of how fast $\ga$ goes to zero. In terms of the representation $\ga(T)=\ga_1 T^\al+\dots$ for $T\to0$  used in \cite{bord10-82-125016}, the dependence on $\ga_1$ drops out for $\al>1$.

It should be mentioned that the above point might occur non-natural when thinking of dielectric bodies described by a fixed $\ep$. As observed in \cite{schw78-115-1}, for the TE mode, the limits $T\to\infty$ and $\ep\to\infty$ (ideal conductor limit) do not commute. Considering the ideal conductor as a zeroth order approximation, the dependence on $1/\ep$ is non--perturbative. However, this is not a good example. One can argue that this model  is not physical since a dielectric with $(\ep-1)$ not decreasing for high frequencies cannot be realized in nature. Indeed, as soon as $(\ep-1)$ decreases like, for example, in the plasma model, this problem disappears.

In the following section, we focus on the TE mode and describe how to formulate the free energy in the Drude model in terms of physical frequencies and how to set up a perturbative expansion. In the third section, we discuss the relation to a representation in terms of imaginary frequencies. The conclusions are drawn in the last section.\\
Throughout this paper we put $\hbar=c=k_{\rm B}=1$.
\section{Free energy in terms of physical frequencies}
We start our discussion of  free energy from first principles, not resorting to the Matsubara formalism. In general, the free energy $\F$ is given by the Gibbs sum,
\be\label{F2}\F=-T\ln \mbox{Tr}\,e^{-\beta H},
\ee
where $\beta=1/T$ is the inverse temperature, $H$ is the Hamilton operator, and the trace goes over the corresponding Hilbert space. For bosonic excitations with one-particle energies resp. frequencies $\om_J$, where $J$ is a generic index numbering the excitations, this representation turns into
\be\label{F3a}
\F=T\sum_J \ln\left(2\sinh\left(\frac{\beta \om_J}{2}\right)\right).
\ee
Also, it can be rewritten in the form
\bea\label{F3}
\F&=&\frac{1}{2}\sum_J\om_J^{1-2s}+T\sum_J\ln\left(1-e^{-\beta\om_J}\right),
\nn \\  &\equiv&E_0+\Delta_T\F,
\eea
where we separated the vacuum energy, $E_0$,   which remains for vanishing temperature. In \Ref{F3},  a regularization parameter $s$ is used. We do not touch the question about the temperature dependence that comes into $E_0$ through the permittivity. We are interested only in the second part, $\Delta_T\F$,  which is the temperature--dependent part of the free energy. Of course, representation \Ref{F2} is not new, see eq. (5.29) in \cite{BKMM} for example. This formula is very general and it holds for any bosonic system. In application to our system, we restrict our consideration to the excitations $\om_J$ of the electromagnetic field. For these, we can write $\Delta_T\F$ in a more specific form,
\bea\label{F4}\Delta_T\F&=&
T\int\frac{d^2k_{||}}{(2\pi)^2} \Bigg[ \sum_j \ln\left(1-e^{-\beta \om_j}\right)
\nn\\
&&+\int_0^\infty\frac{dk_3}{\pi}\ln\left(1-e^{-\beta \om_{k_3}}\right)  \frac{\pa}{\pa k_3}\delta(k_3)  \Bigg],
\eea
where, like in \Ref{F1}, $k_{||}$ is the momentum parallel to the planes and in passing from \Ref{F3} to \Ref{F4}, $\Delta_T\F$ became the energy density per unit area. The sum over $j$ goes over the discrete (for a given $k_{||}$) frequencies $\om_j$ corresponding to the waveguide and evanescent modes. These are all modes whose wave functions decrease in the directions $z\to\pm\infty$. The integral over $k_3$ accounts for the photonic modes, the momenta are related by
\be\label{k1}k=\sqrt{k_{||}^2+k_3^2}
\ee
and
\be\label{d1}\delta(k_3) =\frac{1}{2i}\ln\frac{t(k_3)}{t(-k_3)}
\ee
is the scattering phase shift expressed in terms of the transmission coefficient,
\be\label{t1}t(k_3)=\frac{4 k_3 q}{(k_3+q)^2e^{-i q L}-(k_3-q)^2e^{i q L}},
\ee
of the corresponding one-dimensional scattering problem for the TE polarization. Here, $L$ is the width of the gap. The scattering coefficient for  TM polarization differs by  $q$ in the parentheses to be multiplied by $\ep$. In this setup, $k_3$ is the  momentum of the asymptotic states and
\be\label{13}q=\sqrt{\frac{(1-\ep)k_{||}^2+k_3^2}{\ep}}
\ee
is the momentum of the photons in the gap.
These modes, having a real momentum $k_3$, correspond to photons propagating in the whole space and we call them photonic modes. We denote their frequency by $\om_{k_3}$.

The discrete frequencies $\om_j$ appear for imaginary momenta $k_3=i\kappa_j$ corresponding to the poles of the transmission coefficient. Here, we have
\be k=\sqrt{k_{||}^2-\kappa_j^2}
\ee
and the $\kappa_j$ are solutions of the known equations,
\bea\label{sln1}
\frac{\kappa}{q}&=&\tan \frac{q L}{2}, \quad\mbox{(symmetric solution)}
\nn \\
\frac{\kappa}{q}&=&-\cot \frac{q L}{2}, \quad\mbox{(anti-symmetric solution)}
\eea
with $q$ given by
\be\label{15}q=\sqrt{\frac{(1-\ep)k_{||}^2-\kappa^2}{\ep}}.
\ee
For real $q$ the solutions are waveguide modes, and for imaginary $q$, the solutions are the evanescent waves. The latter appear in the plasma model and for  TM polarization only. The eqs. \Ref{sln1} are for  TE polarization. That for  TM polarization follow by multiplying the $q$ in the denominators by $\ep$.

Now, we turn to the frequencies of these modes. From the Maxwell equations, the frequencies are  solutions of the equation
\be\label{feq}\ep\,\om^2=k^2,
\ee
where $k$, eq.\Ref{k1},
is the absolute value of the three--dimen\-sional spatial momentum.
It is essential to understand that, for a frequency--dependent permittivity, $\ep(\om)$,  eq. \Ref{feq} takes the form $\ep(\om)\om^2=k^2$ and that just this equation determines the physical frequencies.

The frequencies appearing in eq. \Ref{F4} are just solutions of this equation, whereby  $\om_j$ are the frequencies of the evanescent and waveguide modes with  momentum $k=\sqrt{k_{||}^2-\kappa_j^2}$ and $\om_{k_3}$ is the frequency of the photonic modes with $k$ given by eq. \Ref{k1}.

For the Drude model with  permittivity \Ref{epD}, we have to insert $\ep\to\ep(\om)=1-\Om^2/(\om(\om+i\ga))$ in eq. \Ref{feq}, and this equation can be rewritten in the form of a third--order polynomial,
\be\label{third}\left(\om^2-k^2\right)(\om+i\ga)-\om\,\Om^2=0.
\ee
It has three roots, $\om_a(\ga,k)$ ($a=1,2,3$). For not too large $\ga\lesssim 2 \Om$, the first root has $\mbox{Im}(\om_1(\ga,k))<0$ and $\mbox{Re}(\om_1(\ga,k))>0$. For  $\ga=0$, it turns into the positive frequency of the plasma model,
\be\label{fpl}\om_1(0,k)_{|\ga=0}=\sqrt{\Om^2+k^2}.
\ee
The second root has
$\mbox{Im}( \om_2(\ga,k))<0$ 
and $\mbox{Re}(\om_2(\ga,k))<0$, and it turns for $\ga=0$ into the negative frequency of the plasma model. These two roots describe propagating modes that are damped according to  dissipation. Their starting points are
$$\om_{1,2}(\ga,0)=\frac{1}{2}\left(-i\ga\pm\sqrt{4\Om^2-\ga^2}\right).$$
The third root $\om_3(\ga,k)$ is purely imaginary. It starts in $\om_{3}(\ga,0)=0$, and it ends in $\om_{3}(\ga,\infty)=-i\ga$. This root represents the over-damped mode in the Drude model. It has no analogy in the plasma model.
We mention that these roots have analytic expressions, which are, however, not really helpful, and we refrain ourselves from displaying them. We mention only the relation
\be\label{sumom}\sum_{i=1}^3\om_i(\ga,k)=i\ga,
\ee
which holds for these roots.
In application to  free energy \Ref{F4}, we consider $\om_1(\ga,k)$ as the frequency of the physical modes.

\begin{figure}
    \hspace{0cm}\includegraphics[width=14cm]{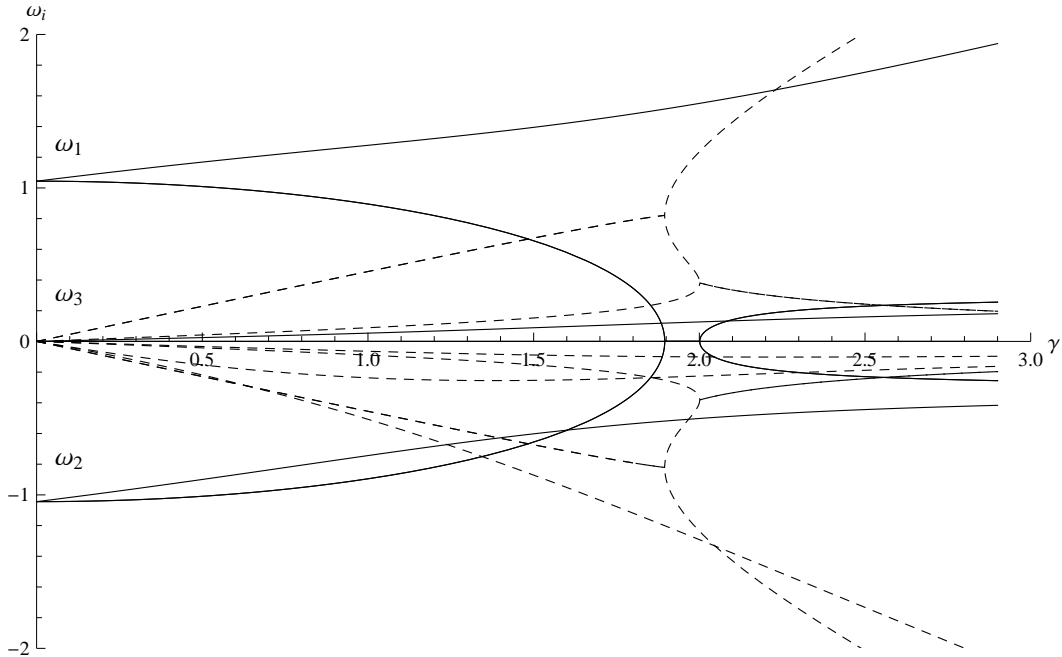}
    \caption{Functions $\om_i(\ga e^{i\alpha},k)$ as functions of $\ga$ for $\alpha=0$, $\alpha=0.2\pi$, $\alpha=\pi$ and $k=0.3$. The solid lines are the real parts, and the dashed lines are the imaginary parts.}
 \label{fgga}
\end{figure}

Now, we return to the point of the perturbative expansion for small $\ga$. From eq. \Ref{third}, it follows that the solutions $\om_i(\ga,k)$ all have power series expansions in $\ga$ with a finite radius of convergence. The first few orders of these expansions are
\bea\label{ompert}
\om_1(\ga,k)&=&\sqrt{\Om^2+k^2} 
-\frac{i\ga}{2}\frac{\Om^2}{\Om^2+k^2}
    +\frac{\ga^2}{8}\frac{(\Om^2+4k^2)\Om^2}{(\Om^2+k^2)^{5/2}}+O(\ga^3),
\nn\\
\om_2(\ga,k)&=&-\sqrt{\Om^2+k^2} 
    -\frac{i\ga}{2}\frac{\Om^2}{\Om^2+k^2}
    -\frac{\ga^2}{8}\frac{(\Om^2+4k^2)\Om^2}{(\Om^2+k^2)^{5/2}}+O(\ga^3),
\nn\\
\om_3(\ga,k)&=& - {i\ga} \frac{k^2}{\Om^2+k^2}+O(\ga^3).
\eea
These are Taylor series expansions with a finite radius of convergence, i.e., these series do converge inside a circle,
$|\gamma|\le\gamma_0$,
in the complex plane of $\gamma$. Radius $\gamma_0$ can be found from the explicit formulas for $\om_i(\ga,k)$. Again, these formulas are quite inconvenient. It is easier and more instructive to consider the real and imaginary parts of $\om_i(\ga e^{i\alpha},k)$ as functions of $\ga$ for several values of $\alpha$ (see Fig. \ref{fgga}). For $\ga\lesssim1$, the curves do not intersect. For $\ga\gtrsim2$, we observe bifurcations. So, we conclude that singularities in the complex $\ga$-plane are all on a finite separation from the origin. Hence, the expansion of each $\om_i(\ga,k)$ in powers of $\ga$ has a finite radius of convergence.

From the above formulas, it is seen that the perturbative expansion requires $\Om^2+k^2\ne0$. However, this is just the frequency in the plasma model \Ref{fpl}, which is known to be never zero. We mention that this holds not only for the photonic modes but also  for the waveguide and evanescent modes too. At once, one can see from  eqs. \Ref{ompert} that the coefficients decrease for $k\to\infty$. This allows the conclusion that the expansion can be made under the sign of the integrals in \Ref{F4}.

We continue with mentioning that all functions of $\om$ entering the free energy, including the solutions of eq. \Ref{sln1} and the phase shift \Ref{d1}, have power series expansions in $\om$. This follows from their explicit form for eq.\Ref{d1}, or from the form of the defining equations \Ref{sln1}. Now, if we insert into these the expansion of $\om_i(\ga,k)$,  we again obtain a power series expansion with a finite radius of convergence.
Finally, we insert the expansion obtained into
the free energy \Ref{F4}. Since the integrations converge for each order of the expansion, we conclude that $\Delta_T\F$ has an expansion in powers of $\ga$ with a finite radius of convergence.

The zeroth order of this expansion is the free energy of the plasma model. The first order is proportional to $i\ga$ and purely imaginary. The second order is proportional to $(i\ga)^2$, and it is real. Higher orders behave accordingly.
So, our conclusion in this section is that the free energy of the first of the considered systems goes to zero for the vanishing relaxation parameter $\ga$. Since we expect the complete system to have the same property, we can conclude that the second part, which we do not calculate,  has this property too.
\section{On the impossibility to turn to imaginary frequencies in the Drude model}
\begin{figure}
   \includegraphics[width=12cm]{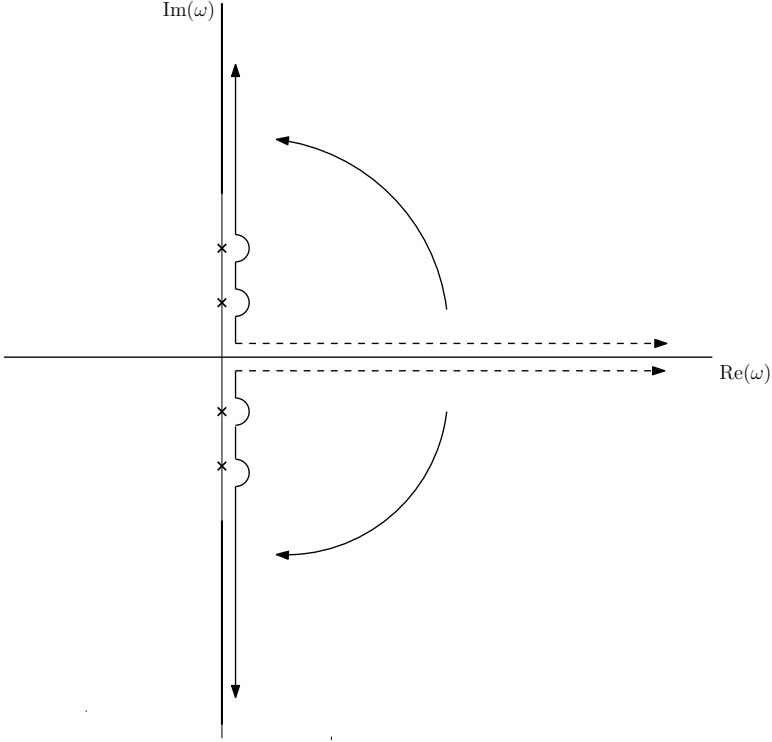}
    \caption{Complex $k_3$-plane for  contour rotation in the beginning of section 3.}
 \label{figk3}
\end{figure}

In this section, we discuss the question about  turning of the momentum integration over $k_3$ in \Ref{F4} to the imaginary axis. The aim is to investigate how to pass from  representation \Ref{F4} in terms of physical frequencies to the standard Matsubara formulation \Ref{F1} which is in terms of imaginary frequencies. We start from reminding this procedure in the plasma model, where it is possible without difficulty. So, we put $\ga=0$ for the moment and look on the structure of the representation \Ref{F4}. We follow the procedure applied in \cite{bord95-28-755}. The integral over $k_3$ can be split into two integrals by writing the logarithm in \Ref{d1} as a difference of two,
\be\label{d2}\delta(k_3)=\frac{1}{2i}\left(\ln t(k_3)-\ln t(-k_3)\right).
\ee
Now, we turn the integration path in the first contribution upwards, $k_3\to i\kappa$, and in the second contribution downwards, $k\to -i\kappa$.  We mention that there are no contributions from the large circles and that there are no singularities the paths might cross. The complex $k_3$-plane is shown in figure \ref{figk3}.
The initial integration path goes along the real axis. It is shown by the dashed lines for the two contributions in eq.\Ref{d2}. On the positive imaginary axis, we have the poles of $t(k_3)$, on the negative imaginary axis, poles from $t(-k_3)$. Further, after the poles, there come the cuts starting from $\pm i\sqrt{\kappa^2-\Om^2-k_{||}^2}$. Now, the integrals from below the cuts have contributions from the poles that just cancel the contributions from the discrete modes, i.e., the first part in the square bracket in \Ref{F4}. After that, the integrals along the cuts remain. Here, the two logarithms in \Ref{d2} become equal -- however, the other logarithms are now different since the frequency becomes imaginary, $\om=i\xi$ in the contribution turned upwards and $\om=-i\xi$ in the other, with
\be\label{22}\xi=\sqrt{\kappa^2-\Om^2-k_{||}^2}.
\ee
For these logarithms, we note
\be\label{23}
\ln\left(1-e^{-\beta\om}\right)=
-i\frac{\beta\xi}{2}+\ln\left|2I\sin\frac{\beta\xi}{2}\right|+f_s(\kappa)
\ee
for $\om=i\kappa$ and
\be\label{24}
\ln\left(1-e^{-\beta\om}\right)=
-i\frac{\beta\xi}{2}-i\pi+\ln\left|2i\sin\frac{\beta\xi}{2}\right|-f_s(\kappa)
\ee
for $\om=-i\kappa$.  Function
\be\label{25}f_s(\kappa)=i\pi\sum_{l=1}^{\infty}\Theta(\kappa-\kappa_l)
\ee
with
\be\label{26}\kappa_l=\sqrt{\xi_l^2+\Om^2+k_{||}^2}
\ee
is a stair function appearing from passing the points $\frac{\beta\xi}{2}=\pi l$ ($l$ integer), where the sinus changes sign. Obviously, $\xi_l$ is the Matsubara frequency \Ref{xil}.  Next, we observe that  term $(-i\beta \xi)$ in the parenthesis just cancels the vacuum energy $E_0$, which was separated in \Ref{F3}. So, we continue with the complete free energy $\F$. Finally, we integrate by parts and come to
\bea\label{F6}\F&=&\int\frac{d k_{||}}{(2\pi)^2}\Bigg\{-\frac12\ln t(i\kappa_0)
-\int_{\sqrt{\Om^2+k_{||}^2}}^{\infty}d\kappa\, \frac{1}{i\pi}\frac{\pa f_s(\kappa)}{\pa\kappa}\ln t(i\kappa)\Bigg\}.
\eea
The first contribution in the curly brackets is the surface term  from the $(-i\pi)$-contribution. Now, the derivative of the stair function gives the sum of delta functions, which allow to carry out the $\kappa$-integration and, as a result, the Matsubara sum occurs,
\be\label{F7}\F=T\int\frac{d k_{||}}{(2\pi)^2}\,{\sum_{\l=0}^\infty}'\ln t(i\kappa_l).
\ee
This expression is known to be a version of the Matsubara sum representation of  free energy in the plasma model. For example, it coincides up to notations with eq. (5.5) in \cite{bord01-353-1}. Also, it coincides with \Ref{F1} up to a contribution that does not depend on  separation $L$ and up to the sum over  polarizations that we did not indicate in this section.

In this way, we have seen for the plasma model how to pass from the representation of the free energy as a sum/integral over real frequencies, i.e, over the physical excitations, to the Matsubara representation. Now, we consider the question on whether this is possible for the Drude model too. We take  representation \Ref{F4} with the frequency $\om_1(\ga,k)$ as that of the physical excitations of the electromagnetic field, which need to be accounted for in the free energy.

So, we start from considering  solution $\om_1(\ga,k)$ defined in the preceding section. When turning $k_3\to i\kappa$, we have to turn the complete spatial momentum
\be\label{29}k=\sqrt{k_{||}^2+k_3^2}\to i\xi\equiv i\sqrt{k_3^2 -\kappa_{||}^2}
\ee
$\om$ depends on.
Therefore, we first investigate $\om_1(\ga,k)$ for $k=\xi e^{i\al}$ under  deformation $\al=0\dots\frac{\pi}{2}$.

\begin{figure}
    \includegraphics[width=15cm]{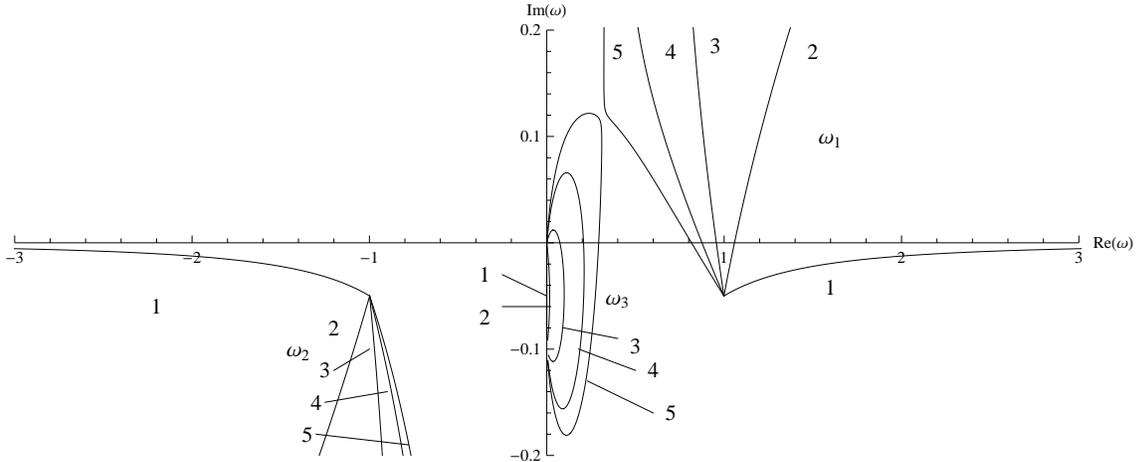}
    \caption{Parametric plots of  functions $\om_a(\xi e^{-i\al})$ in the complex $\om$-plane as a function of $\xi$ for deforming the path.  Curves 1 to 5 correspond to $\al/(\pi/2)=0,\,0.2,\,0.7,\,0.85,\,0.885$, respectively. The curves starting from $\om=0.99-0.05i$ represent the first solutions, $\om_1$, that joining $\om=0$ with $\om=-i$ represent the third solution, $\om_3$, and that in the third quadrant represent the second solution, $\om_2$. The value of the relaxation parameter is $\ga=1$.}
 \label{fg1}
\end{figure}

\begin{figure}
    \hspace{0cm}\includegraphics[width=14cm]{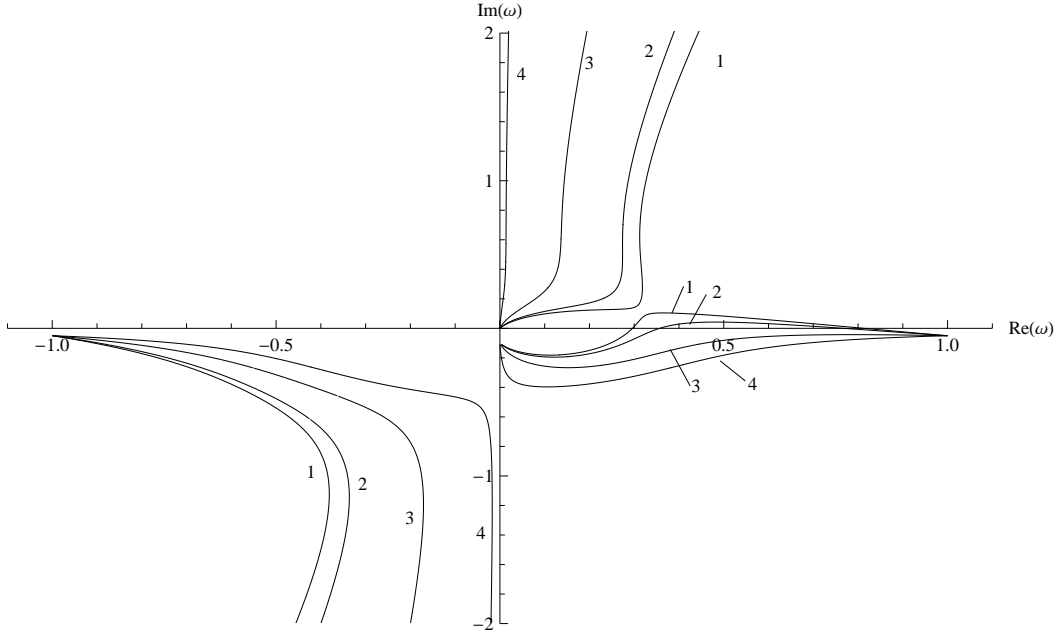}
    \caption{Parametric plots of  functions $\om_a(\xi e^{-i\al})$ in the complex $\om$-plane as a function of $\xi$ for deforming the path.  Curves 1 to 4 correspond to $\al/(\pi/2)=0.886,\,0.9,\,0.95,\,0.995$, respectively.  The value of the relaxation parameter is $\ga=1$.}
 \label{fg2}
\end{figure}

This task is a bit cumbersome, however, it does not pose any principal difficulties. For the  understanding of  deformation, it is necessary to consider all three solutions, $\om_a(\ga,k)$ ($a=1,2,3$), together. These are shown (in arbitrary units) in  figures \ref{fg1} and \ref{fg2} as parametric plots in the complex $\om$-plane as functions of $\xi$ for several values of $\al$. The relaxation parameter is taken to be $\ga=0.1$, and the plasma frequency is put $\Om=1$.
First, we consider $\al\le0.885\frac{\pi}{2}$, which is shown in figure \ref{fg1}. Solutions $\om_{1,2}$ start from the points
\be\label{30}\om_{1,2}(\ga,0)=\frac12\left(-i\ga\pm\sqrt{4\Om^2+\ga^2}\right),
\ee
whereas $\om_3$ starts from $\om_3(\ga,0)=0$. These points stay fixed under the deformation of $\al$. For $\xi\to\infty$,  solutions $\om_{1,2}$ behave like the corresponding frequencies in the plasma model, i.e., $\om_{1,2}(\ga,k\to\infty)=\pm k+\dots$. The third solution takes imaginary values for all $\xi$ while $\al=0$. It goes on the imaginary axis from 0 to $-i\ga$.

Now we increase $\al$. The curves deform and are denoted by  numbers 2 to 5 for increasing $\al$. It is seen that the first two solutions move towards the corresponding parts of the imaginary axis, while the third solution deforms to the right acquiring a positive real part. At some critical value, $\al^*\sim0.8855$ in the given case, the solutions $\om_1$ and $\om_3$ touch. Beyond, for $\al>\al^*$, they detach, but in a changed order. That means that the correspondence between starting points (for $\xi=0$) and end points (for $\xi\to\infty$)  changed.  Hence, a turning of the integration path beyond $\alpha^*$ is impossible. So, we have to conclude that within the Drude model it is not allowed to turn from the physical frequencies to the imaginary ones. Equivalently, it is not possible to come to the Matsubara representation.

It is interesting to see what could be done about this problem.  Indeed, it is possible to turn the contour beyond $\alpha^*$. For this, one needs to include from the beginning the contribution from the over--damped mode, $\om_3(\ga,k)$, in eq. \Ref{F4}. In that case,  for $\al>\al^*$, we have a part of the integration path composed of the former $\om_1$ and $\om_3$ going from $\om_3(\ga,0)$ to $\om_1(\ga,\infty \, e^{i\al})$. These are the curves in the first quadrant in figure \ref{fg2}. For $\al=\frac{\pi}{2}$ the final curve of  deformation goes just along the positive imaginary axis. This is obviously the curve used in the standard Matsubara representation for the Drude model \Ref{F1}. In addition, we have another contribution from the path joining the starting point of $\om_1$ with the end point of $\om_3$. These are the curves in the fourth quadrant. These are complex also for $\al=\frac{\pi}{2}$ and, if accounted for,  deliver a complex contribution to the free energy. In the standard Matsubara representation, this contribution is not present. In this way, the Matsubara representation accounts for a part of the frequencies originating from $\om_1(\ga,k)$ and from $\om_3(\ga,k)$  and disregards another part of these.

In addition, we would like to mention that  solution $\om_3(\ga,k)$ (eddy currents) being inserted into  free energy   \Ref{F4} causes a problem with the convergence of the integrals for large momenta. First, we mention that the convergence in \Ref{F4} results from the decrease in factor $\ln(1-e^{-\beta \om})$ for $\om\to\infty$. Now,  frequency $\om_3(\ga,k)$ is purely imaginary. Hence, the logarithm does not decrease. One is left with the hope that it oscillates sufficiently fast to ensure convergence. However, this is also not the case since $\om_3(\ga,k)$ is bounded, $0\le -i \om_3(\ga,k)\le 1$, for real $k$. Finally, one could argue that the convergence in \Ref{F4} could come from the decrease in the phase shift. However, the  integration over $k_{||}$ remains divergent.

In \cite{intr10-82-032504},
the modes corresponding to  solution $\om_3(\ga,\om)$ (these are the 'eddy' currents), were included in addition to $\om_1(\ga,\om)$ in the sum over modes in \Ref{F3} or \Ref{F4}. However, this alone is also not sufficient to get a real free energy, resp. the standard Matsubara representation, what is the aim of \cite{intr10-82-032504}. Even for $\al=\pi/2$, i.e., at the final deformation of the path, from $\om_3(\ga,\om)$ a part of the path is in the complex plane (curve 4 in Fig. 3). In addition, in \cite{intr10-82-032504}, also the modes corresponding to the solutions $\om_2(\ga,\om)$ are included. In that case, all the imaginary contributions compensate each other and one comes to a real free energy (already on the level of eq.\Ref{F3} or \Ref{F4}). However, we cannot include  mode $\om_2(\ga,\om)$ because in that case, when setting $\ga=0$, we would get the plasma model twice. In addition, there is another problem with the   modes corresponding to  solutions $\om_2(\ga,\om)$. Their negative real part would cause the sum over the eigenvalues $J$  in eq. \Ref{F3} to diverge. Of course, this divergence does not enter the force (it does not depend on  separation $L$, as can be seen in eq. \Ref{F3a}); however, it would produce an unusual temperature--dependent divergence.

In the remaining part of this section, we will bring an argument showing that the Drude model inserted into the standard Matsubara formula has a contribution that is non perturbative in $\ga$. For that, we consider a contour rotation like above, but in the opposite direction. We start from the standard Matsubara representation \Ref{F1} of the free energy in the Drude model. It can be represented by eq.\Ref{F7}, which is the free energy for the plasma model in Matsubara representation, where one has to insert \Ref{epD} at imaginary frequency,
\be\label{31}\ep=1+\frac{\Om^2}{\xi(\xi+\ga)}.
\ee
The corresponding frequency condition, which gives the connection with the imaginary momentum perpendicular to the plane, reads
\be\label{32}-\xi^2=\frac{\Om^2}{1+\ga/\xi}+k_{||}^2-\kappa^2.
\ee
In opposite to the third--order equation \Ref{third} for $\om$, this is a second--order equation for $\kappa$, which can be  solved explicitly,
\be\label{kxi}\kappa(\xi)=\sqrt{\xi^2+\frac{\Om^2}{1+\ga/\xi}+k_{||}^2}\,.
\ee
Now, we go the way back from representation \Ref{F7}, or from eq. \Ref{F1}, to a representation in terms of real frequencies. This can be done by going from eq. \Ref{F7} backwards or by applying the Abel-Plana formula to the sum over $l$ in \Ref{F7}. The latter procedure was used, for example, in \cite{bord10-82-125016}, eqs. (11) and (13). In any case, we have to turn the integration path for $\xi$ towards its imaginary axis, $\xi\to\pm i \om$. Let us follow what happens to $\kappa(\xi)$, eq.\Ref{kxi}.
\begin{figure}
    \hspace{0cm}\includegraphics[width=12cm]{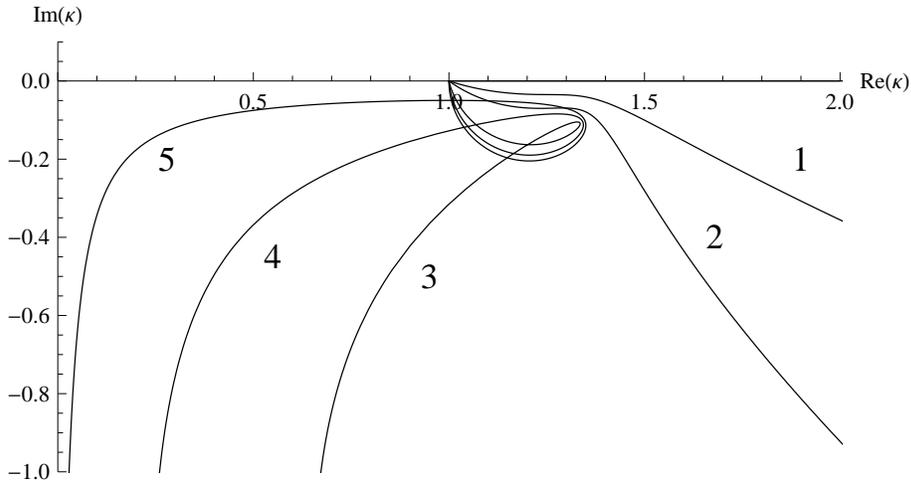}
    \caption{Parametric plot of  function $\kappa(\xi e^{-i\al})$, eq. \Ref{kxi}, in the complex $\kappa$-plane as a function of $\xi$ for deforming the path. Curves 1 to 5 correspond to $\al/(\pi/2)=0.2,\,0.4,\,0.85,\,0.95,\,1$, respectively.}
 \label{fg3}
\end{figure}
We show in figure \ref{fg3} the deformation of  function $\kappa(\om e^{-i\al})$ in the complex $\kappa$-plane for $\al=0\dots\frac{\pi}{2}$. Again, we take values $\ga=0.1$, $k_{||}=1$ and $\Om=1$. For $\al=0$, the initial curve is along the real axis from 1 to $\infty$. For increasing $\al$, the curve goes down into the fourth quadrant. These are curves 1 and 2. Further increasing $\al$, at some critical value of $\al$, the curve develops a cusp and, further, a self-intersection, see curve 3. Increasing $\al$ further (curve 4) until the final value of $\pi/2$ (curve 5) the curve starts with its loop and, in the following course, it comes quite close (for small  $\gamma$) to the real and imaginary axes. It is clear that, in this way, one comes to an integration over real frequencies $\om$, however, one does not come to the frequencies of the Drude model which are the first root of  eq. \Ref{third}. Instead, one comes to a representation with integration over real $\om$ but complex $k_3=i\kappa(-i\om)$. This reflects the fact that we do not consider the complete free energy.

In fact, this representation is not really interesting except for    the appearance of the self-intersection and of the loop since the loop is a contribution which is non-perturbative in $\ga$.
\begin{figure}
    \hspace{0cm}\includegraphics[width=13cm]{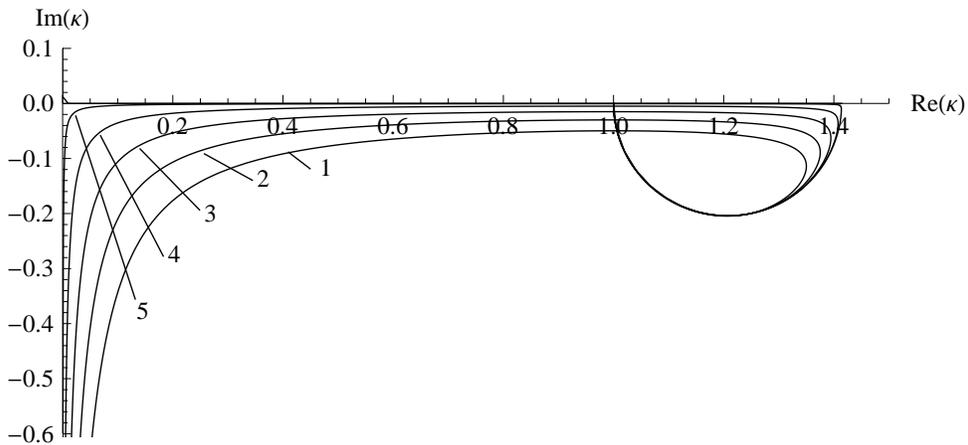}
    \caption{Parametric plot of  function $\kappa(-i\xi)$, eq. \Ref{kxi}, in the complex $\kappa$-plane as a function of $\xi$ for decreasing values of $\ga$.  Curve 1 corresponds to $\ga=0.1$, the curves 2,3,4,5 to $\gamma=0.06,\,0.03,\,0.01,\,0.001$, respectively.}
 \label{fg4}
\end{figure}
We demonstrate this in figure \ref{fg4}, which displays the final path (for $\al=\pi/2$) for decreasing values of the relaxation parameter $\ga$. The path comes closer and closer to the real and imaginary axes, however, the loop remains and it stabilizes at a finite size. The contribution from the parts of the path that are   along the real and imaginary axes deliver, for $\ga=0$,  just the free energy   of the plasma model. Thereby, the contributions from the real axis dropped out. The contribution from the loop does not disappear for any finite $\ga$, no matter how small, and it constitutes just  the non-perturbative contribution.

This non-perturbative contribution can also be calculated analytically  using the methods of \cite{bord10-82-125016}. In fact, in \cite{bord10-82-125016}, limit $T\to0$ was investigated, however, it turned out that  limit $\ga\to0$ (with $T$ fixed) can be treated by the same  method.
We delegate this calculation to the Appendix. The result is
\be\label{nonz}\Delta_T\F^{\rm Drude}(\ga\to0)-\Delta_T\F^{\rm plasma}=\frac{T}{16\pi a^2}f^{\rm D}(0),
\ee
where  function $f^{\rm D}(0)$ given by eq. (55) in \cite{bord10-82-125016} and by eq. \Ref{A10}.
Equation \Ref{nonz} is, of course, just the contribution violating the third law of thermodynamics.
\section{Conclusions}
In the foregoing sections, we considered  free energy using the permittivity of the Drude model. We restricted our consideration to that part of the free energy, which results from the excitations of the electromagnetic field in the presence of the dielectric bodies and, in this way, we ignored the return of the dissipated energy. In this approach, the free energy, or more exactly, a part of it,  is given by eq. \Ref{F4} and it has an imaginary part. We have described the physical modes of the electromagnetic field and identified that which enter \Ref{F4}. Next, we have shown that this free energy has a perturbative expansion for small relaxation parameter $\ga$. Combining  with the discussion in the Introduction, where we argued that the complete free energy must have such an expansion, we can conclude that the second part of the free energy, which we did not consider in this paper, also must have a perturbative expansion in $\ga$. This is the main conclusion from section 2.

In section 3, we discussed the relation between  representation \Ref{F4} of the free energy in terms of physical frequencies and the standard representation (Lifshitz formula) \Ref{F1}, which is in terms of imaginary frequencies. We started from reminding how this transition  goes in the plasma model, where it is possible without any difficulties. Then we tried to do the same in the Drude model. Here, we observed a principal difficulty from the behavior of the integration path under rotation in the complex plane. At some angle of rotation, a kind of bifurcation appears and the path looses its uniqueness. From here, we conclude that it is impossible to pass to the Matsubara representation, at least by contour rotation. Further, we considered the question what happens if ignoring this problem. It turns out that in doing so one comes to the standard Matsubara representation if integrating along a part of an integration path belonging to the over-damped mode in the Drude model and, if, at once, dropping a contribution from some part of the initial integration path belonging to the physical frequencies.

We continued in section 3  with a consideration of the above procedure the other way round. We started from the standard Matsubara representation \Ref{F1} and tried to rotate the integration path back to physical frequencies. Needless to say that this works fine in the plasma model. However, in the Drude model, under rotation, the integration path develops a self-intersection, see figure \ref{fg3}. As a result, the integration path contains a loop in the complex plane, which gives a finite contribution no matter how small the relaxation parameter $\ga$ is made. From here, we have two conclusions. First, and not unexpected from the above discussion, we conclude that one cannot pass from the Matsubara representation to a representation in terms of physical frequencies. Second, we conclude that the Matsubara representation contains a non-perturbative contribution. Also, this is not unexpected after the discussion at the end of the Introduction.

We mention that the procedure, described in the above paragraph, results in the know representation of the Lifshitz formula in terms of real frequencies (see, e.g., eq.  (12.37) in \cite{BKMM}, or eq.\Ref{A3} in the appendix)  for the Drude model too. This is, however, different from the representation in terms of physical frequencies discussed in section 2.

Finally, we consider the question whether in the perturbative approach there is a way to go to the Matsubara representation. In section 2, we have seen that the free energy of the first part of our system has a perturbative expansion for small $\ga$. Thereby, we have seen that  factor $\Om^2+k^2$ appears in the denominator. This is seen explicitly in eq. \Ref{ompert}. From this, the natural conclusion follows that  perturbative expansion is not possible in a representation where this parameter is small. Fortunately, in a representation in terms of physical frequencies, this parameter never becomes small. Next, we mention that the plasma model is the zeroth order approximation for the Drude model. In the plasma model, we have seen in section 3  that, in passing from  physical to imaginary frequencies, i.e., when turning the integration path, this path goes finally along the imaginary axis starting from $\xi=0$. However, in this point, the perturbative expansion is not possible. From here, we draw the conclusion that in the perturbative approach to the Drude model, it is impossible to pass to the standard Matsubara formulation. This backs the above conclusion that \Ref{F1} contains a non-perturbative contribution.

It is also possible to localize the non-perturbative contribution to \Ref{F1} by observing that one can pass from \Ref{F1} to \Ref{F4} if dropping the contribution resulting from integration over the mentioned loop (see Fig. \ref{fg4}). This contribution is complex and it has a finite, nonzero limit for $\ga\to0$. At the end of section 3 we have shown the corresponding analytic expression. This is just the contribution from the loop in the integration path, which causes the trouble with thermodynamics.

From the above, our conclusion is that for the Drude model one does not have a Matsubara representation. The commonly used one \Ref{F1} cannot be correct since it contains a non-perturbative contribution. With eq. \Ref{F4}, one has a part of the correct representation. However, one needs to add the other, still unknown, part that is responsible for the return of the dissipated energy to the electromagnetic field.

We would like to mention two papers, \cite{kupi92-46-2286} and \cite{rosa10-81-033812}, where it was found that the complete system is perturbative in $\gamma$ in the sense we discussed above. In these papers, dissipation is included by coupling of the polarization fields (oscillators constituting the medium)  to a heat bath like in the Huttner-Barnett model. The equations of motion for the quantized heat bath field are solved and inserted into the Maxwell equations. As a result, dissipation and noise polarization appear. Then, the energy is divided into two parts, $W_1$ and $W_2$, in the notations of \cite{rosa10-81-033812}, corresponding to the electromagnetic field with dissipation ($W_1$) and the noise part ($W_2$) which provides the return of energy. That part of the free energy, which we discuss in this paper, just corresponds to $W_1$. Now, in that papers, it is shown that in the final answer one can tend $\gamma$ to zero and arrives just on the original system, without dissipation. In this sense, the models considered in these papers are perturbative in $\gamma$. Regrettably, these papers are of restricted applicability. In \cite{kupi92-46-2286}  a (1+1)-dimensional model is considered, and in \cite{rosa10-81-033812} the model is considered for a homogeneous dielectric medium filling the whole space.

In the present paper, a number of  questions are left unanswered. The most important one is about the derivation of the Lifshitz formula \Ref{F1} in case of dissipation. In  light of the forgoing discussions, we have to expect an inconsistency at some place in the standard derivation. It is clear this  is a quite strong statement that needs further investigation.

Another opened question is about a dielectric with dc conductivity, which is known to have similar problems with the third law of thermodynamics. Here, we expect a similar problem with the Matsubara representation. The point is that for dc conductivity, representation \Ref{F4} is perturbative in  conductivity $\sigma$ as can be checked easily. On the other hand, in the Matsubara representation, the dependence on $\sigma$ is clearly non perturbative. So, the prediction is that the transformation from physical to imaginary frequencies has a loop hole somewhere like that we found  in this paper for the Drude model.

Finally, we return to the discussion in the Introduction of considering the first part of the free energy as an approximation to the complete one. We have seen that this allowed us to get qualitative conclusions about the standard formulation. However, it is not possible to draw quantitative conclusions on how good the approximation is. This means we can calculate the free energy \Ref{F4} numerically. We can do that also perturbatively. The zeroth order is the plasma model, the first order in $\ga$ is purely imaginary and it describes  dissipation. It will be compensated if accounting for the second part of the system describing the return of energy. The second order in $\ga$  is real, and it gives a part of the first correction beyond the plasma model to the complete free energy. However, the second part is still unknown. Moreover, there are no a'priori reasons to expect its contribution to be small as compared to the first one.
Hence,   the only conclusion we have so far, is that the correction is second order in $\ga$. Since in the experimental situations $\ga$ is small (the smallest of all dimensional parameters), this information might be helpful.

\section*{Acknowledgement}
The author benefited from exchange of ideas by the ESF Research Network
CASIMIR, especially on the highly inspiring workshop {\it
Observability and theoretical grounding of thermal Casimir forces}, Trondheim, Jan 26-27, 2011.\\
The author acknowledges very fruitful discussions with G. Klimchitskaya, V. Mostepanenko and, especially, with G. Barton.

\section*{Appendix}
In this appendix we calculate the limit $\ga\to0$ of the free energy of the Drude model and prove eq. \Ref{nonz}. We start from  eq. \Ref{F1} representing the free energy $\F^{\rm Drude}$ in the standard Matsubara representation with  permittivity $\ep^{\rm D}$, \Ref{epD},  of the Drude model inserted. Technically, we follow the calculations in \cite{bord10-82-125016}, eqs. (4)-(16). Applying the Abel-Plana formula we split the free energy,
\be\label{A1}\F^{\rm Drude}=E_0^{\rm Drude}+\Delta_T\F^{\rm Drude},
\ee
into the vacuum energy (this is the Matsubara sum substituted by an integral) and the temperature--dependent part,
\be\label{A2}
\Delta_T\F^{\rm Drude}=\frac{1}{4\pi^2}\int_0^\infty dx\frac{1}{e^{\beta x}-1}
\ i\left(\varphi(ix)-\varphi(-ix)\right).
\ee
Function $\varphi(\xi)$ is given by
\be\label{A3}
\varphi(\xi)=\int_0^\infty dk_{||}k_{||}\,\ln\left(1-r^2e^{-2\eta L}\right)
\ee
for the real argument and $\varphi(\pm ix)$ is its analytic continuation.

As compared to \cite{bord10-82-125016} we changed the notations slightly. So $k$ in \cite{bord10-82-125016} is now $k_{||}$ and $q$ in \cite{bord10-82-125016} is now $\eta$.
The reflection coefficient is given by
\be\label{A4}
r=\frac{\sqrt{\frac{\om_p^2}{1+\ga/\xi}+\eta^2}-\eta}{\sqrt{\frac{\om_p^2}{1+\ga/\xi}+\eta^2}+\eta}
\ee
with
\be\label{A5}       \eta=\sqrt{\xi^2+k_{||}^2}.
\ee
The important step for the following is to divide the integration region in the analytic continuation $\varphi(ix)$ into two parts. The first is $k_{||}\in[0,x]$ and the second is $k_{||}\in[x,\infty)$.

In the first part, we observe that $r$, eq. \Ref{A4}, can be expanded in powers of $\gamma$. This expansion is in fact in powers $\left(\frac{\ga}{ix}\right)^n$. However, since $k_{||}\le x$, there is no singularity in the integrations at least up to the first order ($n$=1). From the zeroth order ($n=0$), we get just the plasma model contribution (note $\eta=i\sqrt{x^2-k_{||}^2}$ here). This can be seen using a procedure similar to that in the beginning of section 3 and the difference is only in the variables used.

In the second contribution,
\be\label{A6}
    \Delta_T\F_2^{\rm Drude}=
    \frac{1}{4\pi^2}\int_0^\infty dx \frac{1}{e^{\beta x}-1}\ i\left(\varphi_2(ix)-\varphi_2(-ix)\right).
\ee
we made in the part of the function $\varphi(ix)$, \Ref{A3}, resulting from $k_{||}\in[x,\infty)$, a change of variable for $\eta=\sqrt{k_{||}^2-x^2}$ (which is real) and denote the result by $\varphi_2(ix)$. It can be written in the form
\be\label{A7}   \varphi_2(ix)=\int_0^\infty d\eta\,\eta\,\ln\left(1-r^2e^{-2\eta L}\right),
\ee
where now
\be\label{A8}   r=
\frac{\sqrt{\frac{\om_p^2}{1+\ga/ix}+\eta^2}-\eta}{\sqrt{\frac{\om_p^2}{1+\ga/ix}+\eta^2}+\eta}
\, .\ee
The integrals in \Ref{A6} with \Ref{A7} inserted do converge. If we put $\ga=0$ in \Ref{A7}, we get formally zero since $\varphi_2(\pm i x)_{|_{\ga=0}}$ is real. In this way, $\Delta_T\F_2^{\rm Drude}$, eq. \Ref{A6}, does not appear in the plasma model at all.

The picture is different if we consider \Ref{A6} for non-zero $\gamma$ and perform  limit $\gamma\to0$. This limit cannot be performed under the sign of the integrals directly. The expansion of the intergrand would start with $\gamma/ix$ and the $x$-integration would linearly diverge for $x\to0$. However, following the equation that goes in \cite{bord10-82-125016} without number between eqs. (51) and (52), one can substitute the $x$-integration for a $\zeta$-integration by $x=\ga\zeta$. One comes to the representation
\be\label{A9}
    \Delta_T\F_2^{\rm Drude}=
    \frac{1}{4\pi^2}\int_0^\infty d\zeta\frac{\ga}{e^{\beta \ga\zeta}-1}\ i\left(\varphi_2(i\ga\zeta)-\varphi_2(-i\ga\zeta)\right).
\ee
From \Ref{A7} and \Ref{A8}, it is seen that $\varphi_2(i\ga\zeta)$ is in fact independent on $\ga$. Now the limit $\ga\to0$ can be performed in \Ref{A9} under the sign of the integral, and we get
\bea\label{A10}
    \Delta_T\F_2^{\rm Drude}(\ga\to0)&=&
    \frac{T}{4\pi^2}\int_0^\infty  \frac{d\zeta }{\zeta}\ i\left(\varphi_2(i\ga\zeta)-\varphi_2(-i\ga\zeta)\right),
  \nn\\  &\equiv&\frac{T}{16\pi^2L^2}\,f^{\rm D}(0),
\eea
which justifies \Ref{nonz}. We mention that the rhs of eq.\Ref{A10} is, up to notations, just eq. (55) in \cite{bord10-82-125016}.

\end{document}